\documentclass[prb,showpacs,twocolumn]{revtex4}

\usepackage{graphicx}

\begin{document}

\title{Cumulant $t$-expansion for strongly correlated fermions}
\author{A. K. Zhuravlev}
\email{zhuravlev@imp.uran.ru} \affiliation{Institute of Metal Physics, 620990 Ekaterinburg, Russia}
\date{\today}

\begin{abstract}
A systematic nonperturbative scheme is implemented to calculate the ground state energy for a wide class of strongly
correlated fermion models. The scheme includes: (a) method of automatic calculations of the cumulants of the model
Hamiltonian; (b) method of the ground state energy calculation from these cumulants using the $t$-expansion proposed by
Horn and Weinstein [Phys. Rev. D \textbf{30}, 1256 (1984)] with new procedure of its extrapolation to $t\rightarrow\infty$.
As an example of application of the method all cumulants up to the 8-th order for spinless fermion model are calculated
exactly, and converging sequences of approximations to the ground state energy are obtained for one-, two- and
three-dimensional versions of the model.

\end{abstract}

\pacs{05.30.Fk, 71.10.Fd, 71.15.-m}

\maketitle

\section{Introduction}

The problem of strongly correlated quantum many-body systems is one of the most complicated in theoretical physics. With
the exception of a few of simplified models this problem cannot be solved analytically, so that one must resort to
numerical methods. But here a researcher is faced with serious difficulties. For example, exact diagonalization runs into
the exponential growth of the Hilbert space dimension with increasing size of the system and therefore is limited to small
clusters, even when using the Lanczos algorithm~\cite{Dagotto1994}. A more sophisticated Density-Matrix Renormalization
Group technique with high-energy states truncation~\cite{DMRG} gives excellent results for the ground state energy of
one-dimensional Fermi systems, but has its own limitation when applied to two- and three-dimensional
cases~\cite{Liang_Pang}. Quantum Monte Carlo method \cite{QMC} can potentially handle larger systems. However, the method
works poorly at low temperatures for fermion systems because of the so called ``minus-sign'' problem~\cite{QMC_minus-sign}.

As an alternative, a certain interest in the construction of regular expansions still exists \cite{Oitmaa}. The attractive
feature of such alternative is the relative simplicity of calculation of terms in the expansion. Unfortunately, the series
expansions in powers of the coupling constants usually diverge~\cite{Dyson}. However, there are regular expansion methods
which are not reduced to power expansion in coupling constant. For example, high-temperature expansion is that one worth to
mention \cite{Phase_Trans_3}.

Another one is the so called $t$-expansion~\cite{HornWeinstein}, which we describe briefly in what follows. Given a
Hamiltonian $\hat{H}$ and an initial state $|\phi_0\rangle$, let us define the moments
\begin{equation}
\mu_m = \langle\phi_0| \hat{H}^m |\phi_0\rangle \label{eq:mu_n}
\end{equation}
($|\phi_0\rangle$ is normalized to unity) and introduce auxiliary function
\begin{equation}
E(t) = \frac{\langle\phi_0| \hat{H}e^{-\hat{H}t} |\phi_0\rangle}{\langle\phi_0| e^{-\hat{H}t} |\phi_0\rangle}
\label{eq:E_t_first}
\end{equation}
which can be written as a power series in the parameter~$t$:
\begin{equation}
E(t) = \sum_{m=0}^\infty \frac{I_{m+1}}{m!}(-t)^m \ ,
\end{equation}
where
\begin{equation}
I_{m+1} = \mu_{m+1} - \sum_{p=0}^{m-1} {m \choose p} I_{p+1} \mu_{m-p} \label{cumulants}
\end{equation}
are the cumulants\cite{Smith} (note that in \cite{Cioslowski,Stubbins,Oitmaa} the values $I_m$ were named ``connected
moments''). Then
\begin{equation}
E_0 = \lim_{t\rightarrow\infty} E(t) \label{eq:E0lim}
\end{equation}
is the minimal eigenvalue of the Schr\"{o}dinger equation
\begin{equation}
\hat{H}|\psi_0\rangle = E_0 |\psi_0\rangle \label{eq:Hpsi}
\end{equation}
provided that $\langle \psi_0 |\phi_0\rangle \neq 0$ (see \cite{HornWeinstein} for proof).

There were attempts to use $t$-expansion in lattice gauge theory \cite{Morningstar}, quantum chromodynamics
\cite{Schreiber}, quantum chemistry \cite{Cioslowski}. In condensed matter physics $t$-expansion was applied to the square
lattice Heisenberg antiferromagnet  \cite{Zheng_Oitmaa_Hamer_1995, Mancini_2005}. For the models of interacting electrons
on a lattice this method is not used mainly for two reasons: 1) it is difficult to calculate cumulants $I_n$ for any
realistic model, 2) it is not easy to calculate the limit (\ref{eq:E0lim}) having the finite number of known cumulants. In
this paper we present a solution to both of these problems and test it for the spinless fermion model which is a typical
example of strongly correlated fermion models.

\section{Calculation of the cumulants}

The Hamiltonian of this model reads
\begin{eqnarray}
\hat{H} &=& \hat{W} + \hat{V}  \ , \nonumber \\
\hat{W} &=& -w\sum_{i>j} c_{i}^{\dagger }c_{j} + c_{j}^{\dagger }c_{i} \ ,  \hat{V} = v\sum_{i>j} c_{i}^{\dagger }c_{i}
c_{j}^{\dagger }c_{j} \label{eq:Ham}
\end{eqnarray}
with $i$ and $j$ being nearest neighbor lattice sites.

The one-dimensional spinless fermion model is equivalent to the exactly solvable spin-$\frac{1}{2}$ {\it XXZ}
model~\cite{Des_Cloizeaux}. Therefore the model~(\ref{eq:Ham}) in one dimension is often used to test new methods of
calculations (see, e.g., \cite{DMRG}). Note that for the half-filled case of the model the metal-insulator transition takes
place with the appearance of the gap in the energy spectrum at $v>2w$.

If the initial wavefunction $|\phi_0\rangle$ has the form
\begin{eqnarray}
|\phi_0\rangle = \prod_{l} c_{l}^\dagger |0\rangle
\end{eqnarray}
($|0\rangle$ is the state without fermions, $l$ runs over some set of the lattice sites) then many-operator average
included in cumulant $I_n$ can be calculated using Wick's pairing technique. We have to

\noindent 1) connect each creation operator $c_i^\dagger$ with one of the annihilation operators $c_j$ with lines by all
the ways possible,

\noindent 2) for each way of the connection assign a term with factor $(-1)^{P}$, where $P$ is the number of connecting
lines intersections,

\noindent 3) replace each connected pair of operators $c^\dagger_i$ and $c_j$ (or $c_i$ and $c^\dagger_j$) by the average
$\langle c^\dagger_i c_j\rangle_0$ (or $\langle c_i c^\dagger_j\rangle_0$), where we introduced the notation $\langle
\ldots\rangle_0 \equiv \langle\phi_0| \ldots |\phi_0\rangle$.

For example, the calculation of 4-operator average is
\begin{eqnarray}
\label{eq:pairing} \langle c^\dagger_i c_j c^\dagger_k c_l \rangle_0 &=& \langle c^\dagger_i c_j\rangle_0\langle
c^\dagger_k
c_l\rangle_0 + \langle c^\dagger_i c_l\rangle_0 \langle c_j c^\dagger_k\rangle_0 \\
&=& n_i \delta_{ij} n_k \delta_{kl} + n_i \delta_{il} (1-n_k) \delta_{jk} \nonumber
\end{eqnarray}
where $n_i=1$ if $i$-th site in $|\phi_0\rangle$ is filled, and $n_i=0$ if the site is an empty one. Thus each average can
be computed easily, but there are too many of them to perform all the calculations manually. To complete the task the
symbolic manipulation computer program was written that performs these calculations.

To be certain let us consider the one-dimensional half-filled case with the initial state $|\phi_0\rangle =
|10101010...10\rangle$. For this state $\hat{V}|\phi_0\rangle = 0$, so that the terms included in $\langle
\hat{H}^n\rangle_0$ which begin with or end with the operator $\hat{V}$ vanish altogether. Therefore one obtains simpler
expressions for the moments $\mu_n$:
\begin{eqnarray}
\label{mu_second}
\mu_1 &=& \langle \hat{H}\rangle_0 = \langle \hat{W}\rangle_0 \ , \nonumber \\
\mu_2 &=& \langle \hat{H}^2\rangle_0 = \langle \hat{W}\hat{W}\rangle_0 \ , \\
\mu_3 &=& \langle \hat{H}^3\rangle_0 = \langle \hat{W}\hat{W}\hat{W}\rangle_0 + \langle \hat{W}\hat{V}\hat{W}\rangle_0 \ ,
\dots \nonumber
\end{eqnarray} Substituting (\ref{mu_second}) into (\ref{cumulants}) we obtain compact expressions for the
cumulants
\begin{eqnarray}
\label{eq:I_n_connect}
I_1 &=& \langle \hat{W}\rangle \ , \nonumber \\
I_2 &=& \langle \hat{W}\hat{W}\rangle_c \ , \\
I_3 &=& \langle \hat{W}\hat{W}\hat{W}\rangle_c + \langle \hat{W}\hat{V}\hat{W}\rangle_c \ , \dots \nonumber
\end{eqnarray}
where the index ``\emph{c}'' means that only connected terms give contribution in pairings like (\ref{eq:pairing}), i.e.
those in which isolated group of operators $\hat{W}$ and $\hat{V}$ are absent.

Substituting the expressions (\ref{eq:Ham}) for $\hat{W}$ and $\hat{V}$ into (\ref{eq:I_n_connect}), using Wick's pairing
technique and performing necessary analytical calculations with the help of the above-mentioned computer program, we obtain
the final expressions for cumulants of the half-filled one-dimensional spinless fermion model:
\begin{eqnarray}
I_1 &=& 0, \ I_2 = w^2N, \ I_3 = w^2vN, \nonumber \\
I_4 &=& (-6w^4 + w^2v^2)N, \nonumber \\
I_5 &=& (-28w^4v + w^2v^3)N , \nonumber \\
I_6 &=& (160w^6 - 86w^4v^2 + w^2v^4)N , \\
I_7 &=& (1704w^6v - 220w^4v^3 + w^2v^5)N , \nonumber \\
I_8 &=& (-9520w^8 + 10736w^6v^2 - 510w^4v^4 + w^2v^6)N   \nonumber \label{eq:I_n_1D}
\end{eqnarray}
where $N$ is the number of lattice sites. The number of cumulants which could be computed is limited by a computer power
only.

\section{Calculation of the limit $E(t\rightarrow\infty)$}

The next step is the calculation of the limit (\ref{eq:E0lim}). In order to calculate this limit one must know all the
cumulants, which is impossible for any real system. All that we know about the function $E(t)$ is its finite power series
\begin{equation}
E(t) = \sum_{m=0}^{M+1} \frac{I_{m+1}}{m!}(-t)^m \label{eq:E_t_M}
\end{equation}
and the following information: 1) the function  $E(t)$ is a monotonically decreasing one since the derivative of
(\ref{eq:E_t_first}) is the negative of the expectation value of the positive operator $(\hat{H}-\langle
\hat{H}\rangle)^2$, i.e. $\frac{dE}{dt}<0$; 2) $E(t)$ rapidly goes to a constant, hence $\frac{dE}{dt}$ goes to zero as $t$
goes to infinity.
\begin{figure}[h]
\includegraphics[width=\columnwidth, angle=0]{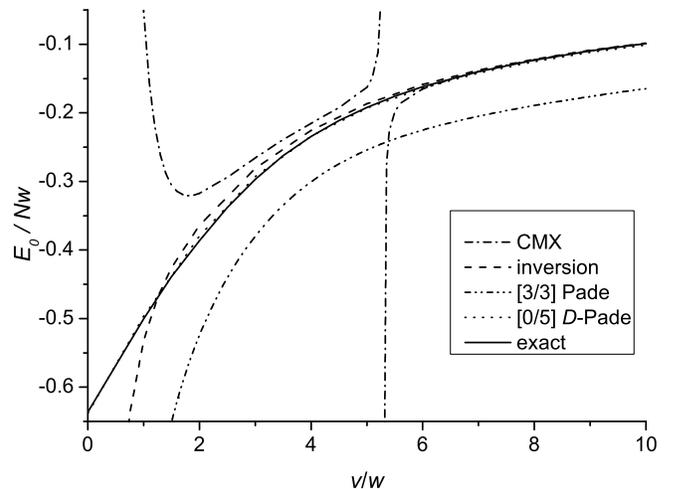}
\caption{The ground state energy density for one-dimensional half-filled spinless fermion model calculated by different
methods using the cumulants 
up to $I_7$} \label{Fig_1Dpade}
\end{figure}
The articles \cite{HornWeinstein,Cioslowski,Stubbins} proposed some ways to calculate the limit (\ref{eq:E0lim}) using the
above information. The Figure \ref{Fig_1Dpade} shows the results of several methods of calculating the limit
(\ref{eq:E0lim}) when the first seven cumulants are known: connected-moments expansion (CMX) by
Cioslowski\cite{Cioslowski}, inversion method by Stubbins\cite{Stubbins}, diagonal [3/3] Pad\'{e} approximant for $E(t)$
and [0/5] $D$-Pad\'{e} approximations used in the pioneering work\cite{HornWeinstein}. The $D$-Pad\'{e} method yields much
more accurate results than others and gives a satisfactory results for the model under consideration in the wide range of
parameters.

It turns out that there is a method which converges more rapidly compared with $D$-Pad\'{e} method. Now we describe it.
The ground-state energy required, $E_0$, can be obtained from
\begin{equation}
\label{eq:int_Eprime}
  \int_0^\infty E'(t)dt = E(\infty) - E(0) = E_0 - I_1 \ ,
\end{equation}
where $E'(t)\equiv \frac{dE(t)}{dt}$, and we have to find the best way to interpolate the function $E'(t)$ between its
known values $E'(0)=-I_2$ and $E'(\infty)=0$.

Let us expand the initial state $|\phi_0\rangle$ in terms of the eigenfunctions of the Hamiltonian as
\begin{equation}
  |\phi_0\rangle = \sum_{n=0}^\infty \sqrt{b_n} |\psi_n\rangle
\end{equation}
with $\hat{H}|\psi_n\rangle = E_n |\psi_n\rangle$. Then the function $E(t)$ can be rewritten as
\begin{equation}
E(t) = \frac{ \int_{E_0}^{E_\mathrm{max}} E e^{-Et}\rho(E)dE }{\int_{E_0}^{E_\mathrm{max}} e^{-Et}\rho(E)dE} \ ,
\label{eq:E_t_DoS}
\end{equation}
where
$\rho(E) = \sum_n b_n \delta(E-E_n)$.
Direct differentiation of (\ref{eq:E_t_DoS}) shows that the asymptotic behavior of the function $E'(t)$ at
$t\rightarrow\infty$ depends on the features of the eigenvalue spectrum. Let us consider the two limiting cases:

\noindent 1) for continuous spectrum with $\rho(E)$ = const $E'(t) \sim -1/t^2$;

\noindent 2) for discrete spectrum $E'(t) \sim -e^{-\Delta t}$, where $\Delta$ is the gap between the ground state energy
$E_0$ and the first excited state energy $E_1$.

\begin{figure}[h]
\includegraphics[width=\columnwidth, angle=0]{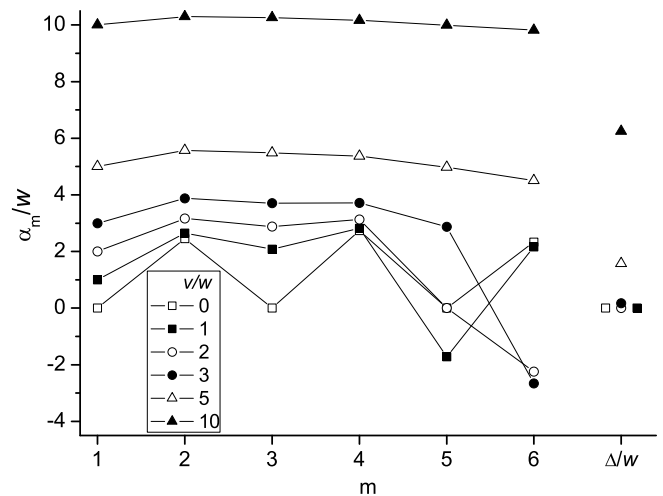}
\caption{The quantity $\alpha_m\equiv$ \textsf{sgn}$(q_m)(m!|q_m|)^{1/m}$ for one-dimensional half-filled spinless fermion
model. The right column shows the exact value of the gap $\Delta$ in the thermodynamical limit\cite{Des_Cloizeaux}}
\label{fig:alpha_1D}
\end{figure}

Now let us consider the function $Q(t)\equiv -I_2/E'(t)$. Its asymptotic behavior must be between $\sim~t^2$ and $\sim
e^{\Delta t}$.
Given the cumulants from $I_1$ to $I_{M+2}$, calculating [0/$M$] Pad\'{e} approximant for $E'(t)$ we obtain the series
expansion for function $Q(t)$ up to order $M$:
\begin{equation}
\label{eq:Q(t)} Q_M(t) = 1 + q_1t + q_2t^2 + \dots + q_Mt^M .
\end{equation}
If the values $q_m$ are close to the coefficients of expansion of the exponential, i.e. there is a good fit to the
dependence
\begin{equation}
\label{eq:q_m} q_m \approx \frac{\alpha^m}{m!} \ (\alpha>0)
\end{equation}
 (see Fig.\ref{fig:alpha_1D}) there is a reason to assume that $Q(t) \sim
e^{\alpha t}$ and take into account the contribution by Kummer's series transformation method\cite{Kummer}. Namely, let us
introduce
\begin{equation}
\label{eq:alpha} \alpha = \frac{1}{M}\sum_{m=1}^M (m!q_m)^{1/m}
\end{equation}
and replace $Q_M(t)$ by $\tilde{Q}_M(t)$:
\begin{equation}
\tilde{Q}_M(t) = Q_M(t) + e^{\alpha t} - \sum_{m=0}^M \frac{\alpha^m}{m!}t^m
\end{equation}
where the first $M$ terms of the series for $\tilde{Q}_M(t)$ coincide with $Q_M(t)$. Then an expression for the approximate
ground state energy is
\begin{equation}
\label{eq:E_DP_0M} E_0(I_{M+2}) = I_1 - \int_0^\infty \frac{I_2}{\tilde{Q}_M(t)}dt ,
\end{equation}

The appearance of negative $q_m$ in (\ref{eq:Q(t)}) should be considered as an indication that the function $Q(t)$ has no
an exponential asymptotics.
Therefore, it has the power asymptotics. Here it is reasonable to use the Pade approximants for $Q(t)$ (or $E'(t)$). For
the integral in (\ref{eq:int_Eprime}) is finite we can use for $E'(t)$ only approximants like [0/$M$],\dots,[$L/M-L$],\dots
where $M-L\geq L+2$. Since $E'(t)$ is always negative we have to control that this property holds for the Pade
approximants. If some approximant has alternating-sign at certain values of $L$ then we will exclude it from consideration.
For each of the remaining proper approximants we calculate the ground state energy $E_0$ according to
(\ref{eq:int_Eprime}).
And if the number of proper approximants is more than one, then we carry out averaging over the energies calculated.

The new extrapolation method described above we call adapted derivative (AD) method.

The results of the ground state energy calculations using the described technique are presented in the Tabl.\ref{tabl:1D}.
As one can see the estimation for the ground state energy converges to its exact value with increasing of the number of the
cumulants known. The new AD-method radically accelerates the convergence rate compared to $D$-Pad\'e approximation for
large $v/w$, that is, where there is a large gap $\Delta$ in the energy spectrum, despite the fact that $\alpha$
(\ref{eq:alpha}) is only a rough estimate for $\Delta$ (see Fig.\ref{fig:alpha_1D}). For $v\lesssim w$ the accuracy of the
AD-method is comparable to the accuracy of the $D$-Pad\'e method. The last case is the most difficult for the method
provided the particle-hole alternating ordered initial state $|\phi_0\rangle$ is chosen.


\begin{table}[htb]
\caption{The sequences of [0/$M$]$D$-Pad\'e and AD approximations to the ground state energy density $E_0/(Nw)$ for
one-dimensional half-filled spinless fermion model using the cumulants up to $I_{M+2}$}
\begin{tabular}{|c|c|c|c|c|c|c|}
\hline $v/w$ & 0 &  2   &   5         &    10 & 20\\
\hline
DP($I_4$) &-0.906900 & -0.553574 & -0.290259 & -0.153778 & -0.078116  \\
DP($I_5$) &-0.906900 & -0.422301 & -0.215831 & -0.113986 & -0.057867  \\
DP($I_6$) &-0.638353 & -0.379567 & -0.198060 & -0.104281 & -0.052850  \\
DP($I_7$) &-0.638353 & -0.379567 & -0.193261 & -0.101028 & -0.051103  \\
DP($I_8$) &-0.625157 & -0.382435 & -0.191984 & -0.099799 & -0.050408  \\
\hline
AD($I_4$) &-0.554850 & -0.348868 & -0.184327 & -0.097832 & -0.049721  \\
AD($I_5$) &-0.615481 & -0.365772 & -0.185731 & -0.098003 & -0.049742  \\
AD($I_6$) &-0.635617 & -0.363360 & -0.187039 & -0.098267 & -0.049780  \\ 
AD($I_7$) &-0.638353 & -0.379567 & -0.188228 & -0.098438 & -0.049802  \\ 
AD($I_8$) &-0.625157 & -0.383881 & -0.189472 & -0.098609 & -0.049824  \\ 
\hline
exact    &-0.636620 & -0.386294 & -0.192014 & -0.099000 & -0.049875  \\
\hline
\end{tabular}
\label{tabl:1D}
\end{table}

\section{Square and simple cubic lattices}
\begin{figure}[htb]
\includegraphics[width=\columnwidth, angle=0]{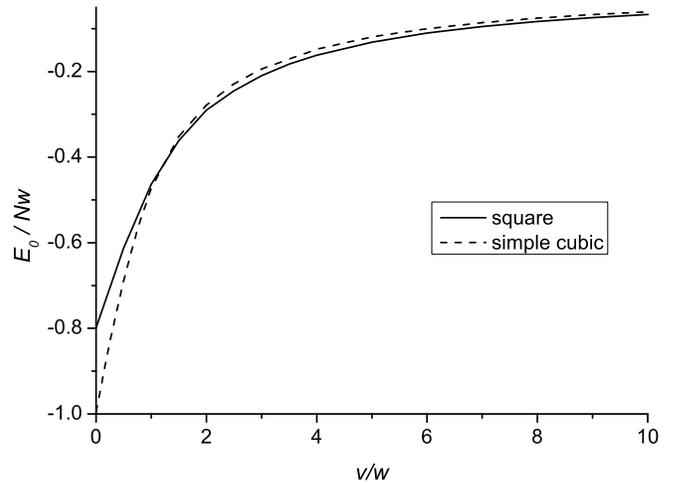}
\caption{The ground state energy density for half-filled spinless fermion model on square and simple cubic lattices}
\label{fig:2D_3D}
\end{figure}

Now let us discuss two- and three-dimensional spinless fermion models, where the exact solutions are not known, except for
a case of non-interacting particles $v=0$. For half-filled spinless fermion model on a square lattice with chessboard
ordered initial state $|\phi_0\rangle$ the cumulants are:
\begin{eqnarray}
I_1 &=& 0 , \ I_2 = 2w^2N , \ I_3 = 6w^2vN , \nonumber \\
I_4 &=& (-36w^4 + 18w^2v^2)N , \nonumber \\
I_5 &=& (-488w^4v + 54w^2v^3)N , \nonumber \\
I_6 &=& (3200w^6 - 4516w^4v^2 + 162w^2v^4)N, \nonumber \\
I_7 &=& (96304w^6v - 35576w^4v^3 + 486w^2v^5)N,  \\
I_8 &=& (-666400w^8 + 1794464w^6v^2 - 257044w^4v^4 \nonumber \\
&+& 1458w^2v^6)N . \nonumber \label{eq:I_n_2D}
\end{eqnarray}
The results for the ground state energy density is presented in Fig.\ref{fig:2D_3D} and Tabl.\ref{tabl:2D}.

\begin{table}[htb]
\caption{The sequences of [0/$M$]$D$-Pad\'e and AD approximations to the ground state energy density $E_0/(Nw)$ for
half-filled spinless fermion model on square lattice using the cumulants up to $I_{M+2}$}
\begin{tabular}{|c|c|c|c|c|c|}
\hline $v/w$ & 0 &   2   &   5         &    10 & 20\\
\hline
DP($I_4$) & -1.047198 & -0.450341 & -0.203638 & -0.103968 & -0.052265  \\
DP($I_5$) & -1.047198 & -0.334617 & -0.150765 & -0.076991 & -0.038708  \\
DP($I_6$) & -0.771728 & -0.304991 & -0.137671 & -0.070281 & -0.035330  \\
DP($I_7$) & -0.771728 & -0.296680 & -0.133138 & -0.067926 & -0.034141  \\
DP($I_8$) & -0.794703 & -0.293364 & -0.131316 & -0.066976 & -0.033662  \\
\hline
AD($I_4$) & -0.874327 & -0.299360 & -0.130763 & -0.066336 & -0.033292  \\ 
AD($I_5$) & -0.971901 & -0.292527 & -0.130079 & -0.066244 & -0.033280  \\ 
AD($I_6$) & -0.767748 & -0.289574 & -0.129868 & -0.066218 & -0.033277  \\ 
AD($I_7$) & -0.771728 & -0.289911 & -0.129840 & -0.066212 & -0.03327588  \\ 
AD($I_8$) & -0.799582 & -0.290245 & -0.129841 & -0.066211 & -0.03327581  \\ 
\hline
exact     & -0.810569 &           &           &           &            \\
\hline
\end{tabular}
\label{tabl:2D}
\end{table}


For half-filled spinless fermion model on the simple cubic lattice with ``three-dimensional chessboard''
ordered initial state $|\phi_0\rangle$ the cumulants are:
\begin{eqnarray}
I_1 &=& 0, \ I_2 = 3w^2N, \ I_3 = 15w^2vN, \nonumber \\
I_4 &=& (-90w^4 + 75w^2v^2)N, \nonumber \\
I_5 &=& (-2052w^4v + 375w^2v^3)N, \nonumber \\
I_6 &=& (14880w^6 - 32058w^4v^2 + 1875w^2v^4)N, \\
I_7 &=& (744600w^6v - 427572w^4v^3 + 9375w^2v^5)N, \nonumber \\
I_8 &=& (-6083280w^8 + 23234064w^6v^2  \nonumber \\
&-& 5240898w^4v^4 + 46875w^2v^6)N . \nonumber \label{eq:I_n_3D}
\end{eqnarray}
The results for the ground state energy density is presented in Fig.\ref{fig:2D_3D} and Tabl.\ref{tabl:3D}.

\begin{table}[htb]
\caption{The sequences of [0/$M$]$D$-Pad\'e and AD approximations to the ground state energy density $E_0/(Nw)$ for
half-filled spinless fermion model on simple cubic lattice using the cumulants up to $I_{M+2}$}
\begin{tabular}{|c|c|c|c|c|c|}
\hline $v/w$ & 0  &  2   &   5         &    10 & 20\\
\hline
DP($I_4$) &-1.216734 & -0.427777 & -0.185303 & -0.093840 & -0.047073  \\
DP($I_5$) &-1.216734 & -0.317453 & -0.137238 & -0.069500 & -0.034864  \\
DP($I_6$) &-0.970781 & -0.290241 & -0.125303 & -0.063438 & -0.031820  \\
DP($I_7$) &-0.970781 & -0.281633 & -0.121129 & -0.061306 & -0.030749  \\
DP($I_8$) &-0.991937 & -0.278155 & -0.119448 & -0.060446 & -0.030317  \\
\hline
AD($I_4$) &-1.015876 & -0.280232 & -0.118592 & -0.059821 & -0.029978  \\ 
AD($I_5$) &-1.129247 & -0.275983 & -0.118229 & -0.059773 & -0.029972  \\ 
AD($I_6$) &-0.963441 & -0.274673 & -0.118121 & -0.059759 & -0.029970   \\ 
AD($I_7$) &-0.970781 & -0.274901 & -0.118099 & -0.059756 & -0.0299692  \\ 
AD($I_8$) &-0.995160 & -0.275095 & -0.118096 & -0.059755 & -0.0299691   \\ 
\hline
exact    &-1.002420 &           &           &           &            \\
\hline
\end{tabular}
\label{tabl:3D}
\end{table}
In both cases, the approximations sequence converge, behaving similarly to the one-dimensional case.

\section{Conclusions}
As we have seen, the present method yields converging sequence of approximations to the ground state energy of a typical
strong-correlated many-fermion model. Seemingly, the sequence of approximations will converge for any Hamiltonian whose
moments (\ref{eq:mu_n}) are finite. The generalization of the method to the case of real electrons with spin is very
simple: one needs to pair only the operators with the same spin indices in the formulas like (\ref{eq:pairing}). Therefore
the method is applicable to real many-electron problems in condensed matter physics and quantum chemistry. The method is of
interest for researchers because it gives a systematic approach to physical problems with strong interaction, which does
not require the smallness of the interaction. The AD-method which we introduce to calculate the limit
$E(t\rightarrow\infty)$ can be useful in the traditional areas of the $t$-expansion application, like the lattice gauge
theory and quantum chromodynamics.

The research was carried out within the state assignment of FASO of Russia (theme ``Electron'' No. 01201463326).


\begin{thebibliography}{*}
\bibitem{Dagotto1994} E. Dagotto, Rev. Mod. Phys. {\bf 66}, 763 (1994).

\bibitem{DMRG}  S. R. White, Phys. Rev. Lett. \textbf{69}, 2863 (1992).

\bibitem{Liang_Pang} S. Liang and H. Pang, Phys. Rev. B \textbf{49}, 9214 (1994).

\bibitem{QMC} J. E. Hirsch, R. L. Sugar, D. J. Scalapino, and R. Blankenbecler,
Phys. Rev. B \textbf{26}, 5033 (1982).

\bibitem{QMC_minus-sign} E. Y. Loh, J. E. Gubernatis, R. T. Scalettar, S. R. White, D. J. Scalapino, and R. L. Sugar,
Phys. Rev. B \textbf{41}, 9301 (1990).

\bibitem{Oitmaa}  J. Oitmaa, C. Hamer and W. Zheng, \textit{Series Expansion Methods For Strongly Interacting Lattice Models}
(Cambridge University Press, Cambridge, 2006).

\bibitem{Dyson}  F. Dyson, Phys. Rev \textbf{85}, 631 (1952).

\bibitem{Phase_Trans_3} \textit{Phase Transitions and Critical Phenomena}, vol.3, edited by C. Domb and M. S. Green
(Academic, London, 1974).

\bibitem{HornWeinstein}  D. Horn and  M. Weinstein, Phys. Rev. D \textbf{30}, 1256 (1984).

\bibitem{Smith}  P. J. Smith, The American Statistician, \textbf{49}, 217 (1995).

\bibitem{Cioslowski}  J. Cioslowski, Phys. Rev. Lett. \textbf{58}, 83 (1987).

\bibitem{Stubbins}  C. Stubbins, Phys. Rev. D \textbf{38}, 1942 (1988).

\bibitem{Morningstar} C. J. Morningstar, Phys. Rev. D \textbf{46}, 824 (1992).

\bibitem{Schreiber} D. Schreiber, Phys. Rev. D \textbf{48}, 5393 (1993).

\bibitem{Zheng_Oitmaa_Hamer_1995} W. Zheng, J. Oitmaa, and C. J. Hamer, Phys. Rev. B \textbf{52}, 10278 (1995).

\bibitem{Mancini_2005} J. D. Mancini, R. K. Murawski,  V. Fessatidis, and S. P. Bowen,
Phys. Rev. B \textbf{72}, 214405 (2005).

\bibitem{Des_Cloizeaux} J. Des Cloizeaux and M. Gaudin, J. Math. Phys. (N.Y.), \textbf{7}, 1384 (1966).

\bibitem{Kummer} \textit{Handbook of Mathematical Functions with Formulas, Graphs,
and Mathematical Tables}, edited by M. Abramowitz and I. A. Stegun (Dover, New York, 1972).

\end{thebibliography}
\end{document}